\documentclass[%
reprint,
superscriptaddress,
amsmath,amssymb,
aps,
]{revtex4-2}

\usepackage{graphicx}% Include figure files
\usepackage{dcolumn}% Align table columns on decimal point
\usepackage{bm}% bold math

\usepackage{booktabs, array, mathptmx, float, tabularx, booktabs, lipsum, amsmath,multirow}
\usepackage{siunitx, xcolor}
\usepackage[colorlinks,linkcolor=blue,anchorcolor=blue,citecolor=blue]{hyperref}

\begin{document}
%\title{Emergent 3$d$-Electron Heavy Fermions at the Narrow Band Close to Mott State in LiV$_2$O$_4$ Thin Films}
\title{Correlation-driven 3$d$ Heavy Fermion behavior in LiV$_2$O$_4$}

\author{Min-Yi-Nan Lei}
\author{Z. H. Chen}
\author{H. T. Wang}
\author{Y. Fan}
\author{N. Guo}
\author{T. X. Jiang}
\affiliation{
Laboratory of Advanced Materials, State Key Laboratory of Surface Physics,
and Department of Physics, Fudan University, Shanghai 200438, China
}%
\author{Yanwei Cao}
\affiliation{
Ningbo Institute of Materials Technology and Engineering, Chinese Academy of Sciences, Ningbo 315201, People's Republic of China
}%

\author{T. Zhang}
\affiliation{
Laboratory of Advanced Materials, State Key Laboratory of Surface Physics,
and Department of Physics, Fudan University, Shanghai 200438, China
}%
\affiliation{
Hefei National Laboratory; Hefei 230088, China
}%
\affiliation{
Shanghai Research Center for Quantum Sciences, Shanghai 201315, People's Republic of China
}
\author{Rui Peng}
\email{pengrui@fudan.edu.cn}
\affiliation{
Laboratory of Advanced Materials, State Key Laboratory of Surface Physics,
and Department of Physics, Fudan University, Shanghai 200438, China
}%
\affiliation{
Shanghai Research Center for Quantum Sciences, Shanghai 201315, People's Republic of China
}%

\author{Haichao Xu}
\email{xuhaichao@fudan.edu.cn}
\affiliation{
Laboratory of Advanced Materials, State Key Laboratory of Surface Physics,
and Department of Physics, Fudan University, Shanghai 200438, China
}%
\affiliation{
Shanghai Research Center for Quantum Sciences, Shanghai 201315, People's Republic of China
}%

\begin{abstract}

LiV$_2$O$_4$ is a spinel-structured compound that stands out as the first known 3$d$-electron system exhibiting typical heavy fermion behavior. A central question is how such strong mass renormalization emerges in the absence of $f$-electrons. In this work, we investigate the three-dimensional electronic structure of LiV$_2$O$_4$ thin films using angle-resolved photoemission spectroscopy (ARPES). We identify that an electron-like flat band is derived from $a_{1g}$ orbitals, along with a highly dispersive $e'_g$ band strongly coupled with phonons.  The overall agreement with dynamical mean-field theory (DMFT) calculations highlights the essential role of inter-orbital Hund’s coupling in reducing the $a_{1g}$ bandwidth to 25~meV, approaching a Mott state. Notably, we find that heavy-fermion behavior arises from additional renormalization at the $a_{1g}$ band near the Fermi level, likely driven by many-body interactions at energy scales down to a few meV and potentially linked to geometric frustration inherent to the spinel lattice. These results provide crucial insights into the origin of the heavy fermion behavior in 3$d$-electron systems.
\end{abstract}

\maketitle
$1.~~Introduction.~$  ~~Heavy fermion materials have long attracted interest in condensed matter physics due to the correlation-driven exotic quantum states, such as unconventional superconductivity and complex magnetic orders ~\cite{white2015unconventional,stewart2017unconventional,machida1989unconventional,RevModPhys.56.755,bauer2004heavy,schroder2000onset}. In typical heavy fermion systems, such as CeCoIn$_5$~\cite{allan2013imaging} and CeCu$_2$Si$_2$~\cite{PhysRevLett.43.1892}, strong electronic correlations  have typically been attributed to the hybridization between localized $f$-orbital electrons and itinerant conduction electrons~\cite{FULDE19881}. 
This hybridization leads to a substantial enhancement of the quasiparticle effective mass, with values ranging from 100 to 1000 times the free electron mass ($m_\mathrm{e}$), as inferred from specific heat measurements~\cite{RevModPhys.56.755,shimoyamada2006heavy}. 
Remarkably, the 3$d$ transition-metal oxide LiV$_2$O$_4$ exhibits a similarly large Sommerfeld coefficient  ($\gamma\simeq 0.42~\mathrm{J/mol~K^2}$), corresponding to an effective mass of approximately 180~$m_\mathrm{e}$. This value rivals those found in typical $f$-electron heavy fermion compounds, despite the absence of $f$-electrons in LiV$_2$O$_4$~\cite{kondo1997liv,JOHNSTON200021}, which presents a longstanding puzzle regarding the microscopic origin of its heavy fermion behavior.

Various scenarios have been proposed to explain the microscopic origin of heavy fermion behavior in LiV$_2$O$_4$. In the spinel structure of this compound, the crystal field splits the V 3$d$ $t_{2g}$ orbitals into a narrow $a_{1g}$ band and broader $e'_g$ bands~\cite{anisimov1999electronic,nekrasov2003orbital}. One proposed mechanism involves hybridization between the more localized $a_{1g}$ states and the itinerant $e'_g$ states, leading to mass renormalization analogous to the $c$-$f$ hybridization~\cite{anisimov1999electronic}. Alternatively, dynamical mean-field theory (DMFT) calculations suggest that LiV$_2$O$_4$ is best described as a lightly hole-doped Mott insulator in the $a_{1g}$ orbital~\cite{arita2007doped}. In this picture, the enhanced quasiparticle mass originates not from hybridization, but from the proximity to a Mott-Hubbard transition.
Hund’s coupling has also been identified as a key factor in driving orbital differentiation and pushing the system toward an orbital-selective Mott-Hubbard transition~\cite{grundner2024liv2o4hundassistedorbitalselectivemottness}.
The geometrical frustration inherent to the spinel lattice [Fig.~1(a)], where V atoms occupy a pyrochlore sublattice, can promote strong spin fluctuations, which may additionally contribute to the heavy effective mass~\cite{knox2013local, lee2017magnetic}. 
Despite these theoretical proposals, a direct experimental determination of the three-dimensional electronic structure near the Fermi level remains lacking. Such information is crucial for elucidating the mechanisms underlying heavy fermion behavior in LiV$_2$O$_4$. Previous photoemission studies have been limited to angle-integrated spectra, partly due to the difficulty of preparing high-quality surfaces in this three-dimensional correlated oxide~\cite{shimoyamada2006heavy}.

Here, we reveal the three-dimensional electronic structure of LiV$_2$O$_4$ by combining film growth with angle-resolved photoemission spectroscopy (ARPES). Our experimental results show overall agreement with DMFT calculations, supporting the pivotal role of inter-orbital Hund’s coupling in driving the $a_{1g}$ orbitals towards a Mott state.The measured dispersion of the flat $a_{1g}$ band deviates from a parabolic form and instead indicates an additional renormalization on a scale of a few meV near the Fermi level. This additional renormalization suggests that the coupling between electrons and low-energy excitations on the few-meV scale, possibly arising from geometric frustration, also plays important roles in the formation of the heavy fermion state in  LiV$_2$O$_4$.

$2.~~Results.$~~Atomically flat LiV$_2$O$_4$ thin films were grown on Nb:SrTiO$_3$(111) substrates for ARPES studies~[Fig.~1(b)]. We fabricated films with a thickness of 72 nm by ablating ceramic targets with a KrF excimer laser ($\lambda = 248$ nm, repetition rate 10 Hz, fluence 0.6 J/cm$^2$) under an argon partial pressure of $1\times 10^{-4}$ mbar at a substrate temperature of 450$^\circ$C. By optimizing the Li/V stoichiometry of the target, we successfully obtained high-quality LiV$_2$O$_4$ films. The XRD results indicate that the films are fully relaxed, with lattice constants consistent with those of bulk LiV$_2$O$_4$ (Appendix A). The reflection high-energy electron diffraction (RHEED) pattern, shown in Fig.~1(c), reveals well-defined two-dimensional streaks, indicating a smooth and ordered surface. The low-energy electron diffraction (LEED) pattern, shown in Fig.~1(d), exhibits sharp spots consistent with the threefold symmetry of the LiV$_2$O$_4$ crystal along the [111] direction. Scanning tunneling microscopy (STM) images show an atomically smooth surface~[Fig.~1(e)]. The details of the film growth process are provided in Appendix A. Temperature-dependent resistivity measurements were performed using a Physical Property Measurement System (PPMS)~[Fig.~1(f)]. At high temperatures, the films exhibit metallic behavior, with resistance decreasing upon cooling. As the temperature is lowered below $T^* \approx$~30~K, the differential resistivity undergoes a sudden change, which signifies a crossover into enhanced electronic coherence \cite{urano2000liv,jonsson2007correlation}. This characteristic temperature is consistent with previous resistivity measurements on bulk crystals~\cite{kondo1997liv, matsushita2005flux} and high-quality thin films~\cite{niemann2023crystallization,PhysRevB.104.245104}.

Most ARPES data were collected at beamline 03U of the SSRF. The data in Figs.~4(l) and (m) were obtained using the He lamp ARPES system at Fudan University, and the data in Appendix B were collected at beamline QMSC of the Canadian Light Source. Both beamlines and the Fudan University ARPES system are equipped with Scienta DA30 electron analyzers. The overall energy resolution was set to 13~meV at Beamline 03U of the SSRF, 10~meV at the Fudan University He lamp ARPES system, and 20~meV at Beamline QMSC of the Canadian Light Source. The angular resolution was $0.1^{\circ}$. ARPES measurements were conducted under an ultrahigh vacuum of $5~\times~10^{-11}$~mbar. At Fudan University, fresh films grown by pulsed laser deposition (PLD) were transferred directly into the ARPES system via an interconnected ultrahigh-vacuum system. The thin films were transferred to the Shanghai Synchrotron Radiation Facility (SSRF) for ARPES measurements in a vacuum suitcase ($1 \times 10^{-10}$~mbar), and the transfer was completed within 12 hours. To minimize surface contamination, the samples were annealed at $300^\circ\mathrm{C}$ for 30 minutes before measurements. A 30-nm-thick amorphous Se layer was deposited as a protective capping for transfer to Beamline QMSC at the Canadian Light Source. Before ARPES measurements, the samples were annealed in ultrahigh vacuum at $500^\circ\mathrm{C}$ for 2 hours to remove the Se layer. The electronic structures of LiV$_2$O$_4$ obtained using different transfer methods are consistent with each other.

\begin{figure}[th]
	\centering
	\includegraphics[width=78mm]{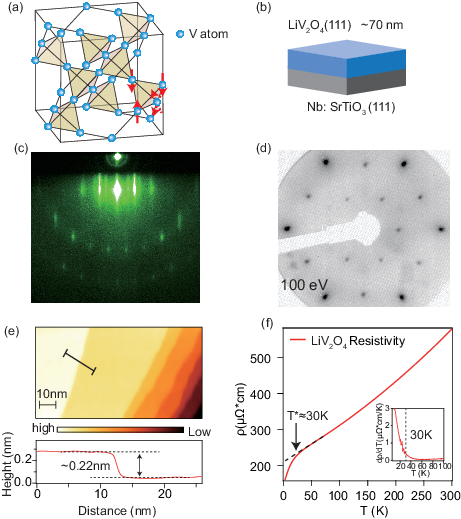}
\caption{
(\textbf{a}) V sublattice in the LiV$_2$O$_4$ crystal. The arrows illustrate the geometric frustration in the corner-sharing V-based tetrahedra. 
(\textbf{b}) Schematic illustration of the LiV$_2$O$_4$ film grown on Nb:SrTiO$_3$ substrate.
(\textbf{c}) RHEED pattern of LiV$_2$O$_4$ thin film. Electron beam was incident along the [11$\bar{2}$] direction.
(\textbf{d}) LEED pattern of LiV$_2$O$_4$ thin film. 
(\textbf{e}) STM topographic image of LiV$_2$O$_4$ thin film surface ($V_\mathrm{b}$ = 1~V and $I$ = 10~pA).
(\textbf{f}) Temperature-dependent resistivity of LiV$_2$O$_4$ thin film. The black dashed line represents the linear extrapolation from the temperature range of 50-100~K. $T^*$ is defined as the temperature at which the resistance curve deviates from the linear extrapolation. The inset shows the derivative of resistance with respect to temperature.
}
	\label{fig:figure212}
\end{figure}

\begin{figure*}[thp]
	\centering
	\includegraphics[width=176mm]{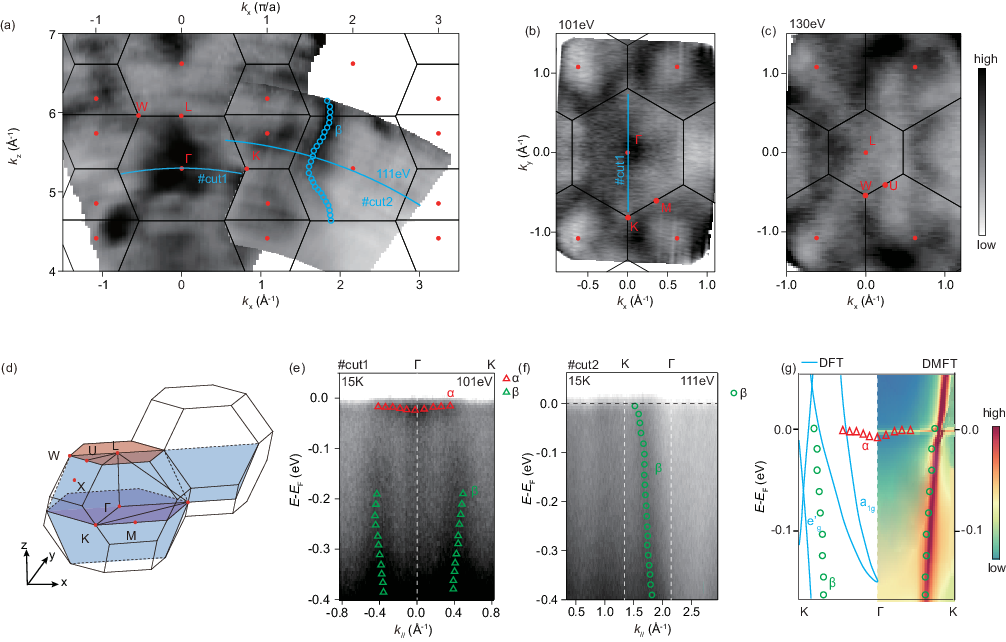}
\caption{ 
(\textbf{a}) Photoemission intensity in the $\Gamma$LW plane, measured using 63~eV to 147~eV photons at 15~K, integrated over $E_\mathrm{F}$ $\pm$ 30 meV.The inner potential was estimated as 15 eV. \emph{$k_\mathrm{F}$}'s (blue open circles) were determined by the local maxima. 
(\textbf{b}) Photoemission intensity in the $\Gamma$KM plane integrated over \emph{$E_\mathrm{F}$} $\pm$ 60~meV, collected with 101~eV photons at 15~K. 
(\textbf{c}) Photoemission intensity in the LWU plane integrated over \emph{$E_\mathrm{F}$} $\pm$ 30~meV, collected with 130~eV photons at 15~K. 
(\textbf{d}) Sketch of the three-dimensional BZ. The momentum regions sampled in panels (a), (b), and (c) are indicated in blue, purple, and magenta, respectively. 
(\textbf{e}) Photoemission spectra along $\Gamma$-K direction (cut~\#1). The $\alpha$ and $\beta$ bands are indicated by triangles determined from the local maxima of MDCs and EDCs. (\textbf{f}) Photoemission spectra of cut~\#2 illustrated in the panel (a). The $\beta$ band is indicated by circles determined from the local maxima of MDCs. 
(\textbf{g}) Comparison of DFT and DMFT  calculations (adapted from Ref.~\onlinecite{grundner2024liv2o4hundassistedorbitalselectivemottness}) with dispersions extracted from photoemission spectroscopy.  
}
	\label{fig:figure2}
\end{figure*}

We measured the electronic structure of LiV\textsubscript{2}O\textsubscript{4} by ARPES. The Fermi surface measured in the $k_x$-$k_z$ plane exhibits strong intensity near the $\Gamma$ point, located around $k_x=0$~[Fig.~2(a)]. At $k_x = 2 \pi/a$, the intensity near the $\Gamma$ point is suppressed due to matrix element effects. Consequently, the intensity of the $\beta$ band Fermi surface, which shows a $k_z$ dispersion consistent with the periodicity of the Brillouin zone, becomes dominant. In both the $\Gamma$KM and LWU planes, the photoemission intensity follows the threefold symmetry of the LiV$_2$O$_4$ crystal along the [111] direction~[Figs.~2(b) and 2(c)].

Along the $\Gamma$–K cut, ARPES resolves two distinct bands: a weakly dispersive $\alpha$ band with relatively flat-band characteristics, and a highly dispersive $\beta$ band [Fig.~2(e)]. Near the Fermi level, the flat $\alpha$ band intensity dominates, while the dispersion of the $\beta$ band is obscured. 
Along cut~\#2 [Fig.~2(f)], the $\alpha$ band is suppressed by matrix element effects, while the $\beta$ band dispersion is better resolved. 
Figure~2(g) compares experimental band dispersions directly with theoretical calculations. The results show a deviation from the DFT calculations but are consistent with DMFT calculations that incorporate electron-electron interactions and Hund's coupling~\cite{grundner2024liv2o4hundassistedorbitalselectivemottness}. Theoretical calculations indicate that the flat $\alpha$ band primarily originates from the $a_{1g}$ orbital, while the dispersive $\beta$ band is predominantly derived from the $e'_{g}$ orbital~\cite{grundner2024liv2o4hundassistedorbitalselectivemottness}. The differences in orbital symmetry lead to different polarization sensitivity in ARPES spectra (see Appendix B). 

%\section*{Electron-Boson Coupling at Dispersive $\beta$ Band}
Higher-resolution spectra reveal a double kink feature in the $\beta$ band [Fig.~3(a)]. The appearance of the kink in the band dispersion is a characteristic signature of boson-electron coupling. By fitting the band dispersion, we identify significant slope changes at binding energies of approximately 75~meV and 150~meV~[Fig.~3(b)]. In the MDCs, we observe increases in peak widths at these two characteristic energies as well~[Fig.~3(c)].
To further quantify these features, we extract the imaginary part of the quasiparticle self-energy (Im$\Sigma$) from the full-width at half-maximum (FWHM) of the MDC peaks [Fig.~3(e)]. The real part (Re$\Sigma$) is estimated from the difference between the experimentally observed dispersion and a bare band dispersion derived from the large-scale spectra~\cite{tang2022antinodal}.  Furthermore, Re$\Sigma$ and Im$\Sigma$ satisfy the Kramers-Kronig relation~[Fig.~3(d)], which imposes additional constraints on the choice of the bare band dispersion. The band renormalization induced by electronic correlations can be estimated from the renormalized Fermi velocities, derived from the dispersion. The renormalized Fermi velocity $v_\mathrm{F}$~=~1.19~eV$\cdot$\AA~is obtained from the slope between the lower-energy kink and $E_\mathrm{F}$, while the bare velocity $v_\mathrm{F}^0$ is determined by the slope over a broader energy range [Fig.~3(b)]. We calculate a renormalization factor $Z = v_\mathrm{F}^0/v_\mathrm{F} = 3.4$ along the $\Gamma$–K direction~\cite{hofmann2009electron}.

Phonon modes near 75~meV have previously been observed in optical reflectivity spectra of several spinel compounds, including MgTi$_2$O$_4$, ZnCr$_2$O$_4$, and NiFe$_2$O$_4$~\cite{PhysRevLett.94.137202, lutz1991lattice}. H. Takagi $et~al.$ suggested that comparable phonon energies are likely present in LiV$_2$O$_4$ due to its structural similarity to these materials~\cite{jonsson2007correlation}. If this is the case, the matching energy scale between the observed kink features and phonon modes suggests that electron-phonon coupling is likely responsible for the 75~meV kink in LiV$_2$O$_4$. 
Future phonon measurements or calculations on LiV$_2$O$_4$ considering the experimental flat band would be invaluable for quantifying the phononic contributions.

Regarding the origin of the 150~meV kink, neutron scattering experiments found no spin-related excitation modes near this energy~\cite{lee2001spin}. The highest-energy phonon modes in LiTi$_2$O$_4$ are below 100~meV~\cite{he2017anisotropic}. Given that high-energy vibrational modes in both LiV$_2$O$_4$ and LiTi$_2$O$_4$ primarily originate from O atom vibrations, phonon modes above~100 meV are not expected in LiV$_2$O$_4$~\cite{ashcroft1976solid}. Thus, we rule out the possibility that the 150~meV kink originates from electron coupling with a single bosonic quasiparticle. 
Notably, DMFT calculations reveal an incoherent $a_{1g}$ spectral weight that approaches $\beta$ at approximately 150~meV below the Fermi level~[Fig.~2(g)], which may influence the dispersion and MDC peak width of the $\beta$ band.
Alternatively, the observed step-like structure of the imaginary part may suggest  two-phonon scattering. 
Specifically, the self-energy exhibits a 68~meV increase from the Fermi level to the first plateau, followed by an additional 34~meV increase to a second plateau [Fig.~3(e)].
The relative magnitude is consistent with previous theoretical studies on weakly polar semiconductors, where two-phonon scattering, despite being a higher-order process, can have comparable strength to one-phonon scattering due to richer scattering channels \cite{lee2020ab}. 
In particular, the phonon at 75~meV in LiTi$_2$O$_4$ corresponds to a longitudinal optical mode, which is known to enable higher-order processes \cite{frohlich1954electrons, vogl1976microscopic}, as evidenced in FeSe/SrTiO$_3$ heterostructures~\cite{liu2021high}. A similar mechanism may also be operative in LiV$_2$O$_4$, where multi-phonon interactions involving longitudinal optical modes could give rise to the 150~meV kink.

\begin{figure}[h]
	\centering
	\includegraphics[width=86mm]{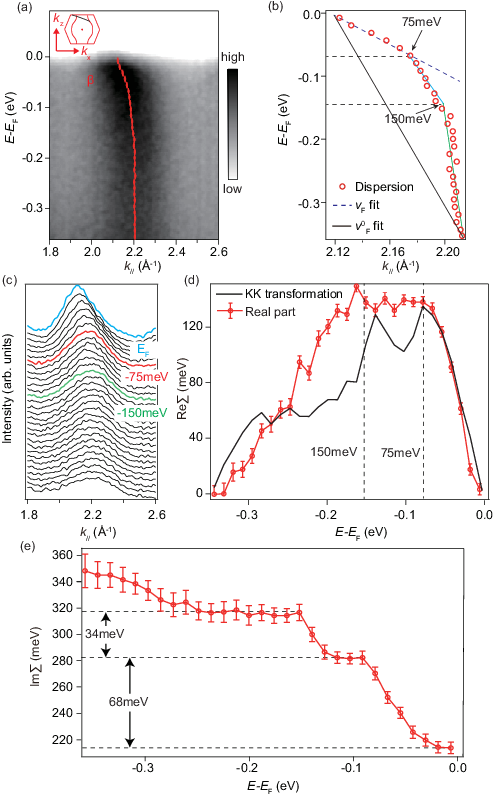}
\caption{
(\textbf{a}) ARPES spectra were acquired using 136 eV photons at 15 K along the cut illustrated in the inset. The overlaid plot represents the dispersion of the $\beta$ band from fitting the MDCs.  
(\textbf{b}) Dispersion of the $\beta$ band in panel (a) (circles), along with the assumed bare band dispersion and fitted Fermi velocity $v_{\mathrm{F}}^0$ (solid line), and the renormalized low-energy dispersion with fitted $v_{\mathrm{F}}$ (dashed line).
 %Dispersion of the $\beta$ band (circles), the assumed bare band dispersion, and $v_{\mathrm{F}}^0$ fit (solid line), ($E-E_\mathrm{F}<$ -300~meV, $v_\mathrm{F}^\mathrm{0}$ fit, solid line) and the renormalized low energy ($E-E_\mathrm{F}>$ -75~meV, $v_\mathrm{F}$ fit, dashed line) dispersion and $v_{\mathrm{F}}$ fit (dashed line).  Dispersion of the $\beta$ band band in panel (a) (circles), the assumed bare band dispersion and $v_{\mathrm{F}}^0$ fit (solid line),  and the renormalized low energy  dispersion and $v_{\mathrm{F}}$ fit (dashed line).
(\textbf{c}) MDCs of the spectra in panel (a). The blue, red, and green MDCs represent energies of $E_\mathrm{F}$, $E_\mathrm{F}$ - 75~meV, and $E_\mathrm{F}$ - 150~meV, respectively.
(\textbf{d}) Real part of the self-energy Re$\Sigma$ (red dots and curve) from the difference between the extracted dispersion and assumed bare band.
(\textbf{e}) Imaginary part of the self-energy Im$\Sigma$ estimated from the full width at half maximum of the MDCs.  The Kramers-Kronig transformation of  Im$\Sigma$ is plotted in panel (d) (black curve) for comparison with Re$\Sigma$.  
}
	\label{fig:figure212}
\end{figure}

\begin{figure*}[thp]
	\centering
	\includegraphics[width=176mm]{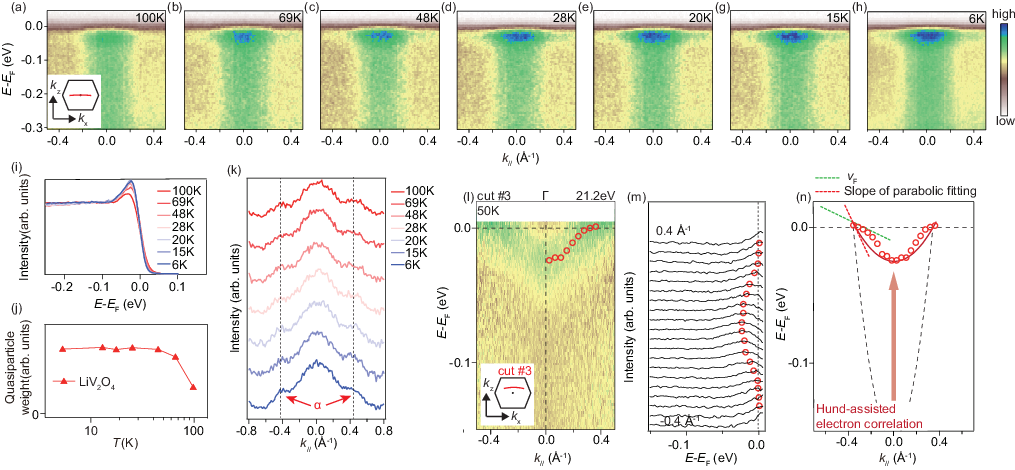}
\caption{
(\textbf{a})-(\textbf{h}) Temperature-dependent photoemission spectra
along $\Gamma$-M direction collected with 101~eV photons. The measurements were performed using linearly horizontal (LH) polarized photons.
(\textbf{i}) Temperature-dependent EDCs integrated over the momentum window [-0.2~\AA$^{-1}$, 0.2~\AA$^{-1}$] around the $\Gamma$ point.
 (\textbf{j}) Flat band weight of LiV$_2$O$_4$ obtained by integrating the spectral weight of the EDCs over the window of [$E_\mathrm{F}$-0.05~eV, $E_\mathrm{F}$+0.05~eV]. The procedure of extracting the flat band weight are explained in Appendix E.
 (\textbf{k}) MDCs integrated over $E_\mathrm{F}\pm5\mathrm{meV}$ for the data in panel (a)-(h). The red arrows indicate the Fermi crossings of $\alpha$ band.
(\textbf{l}) Photoemission spectra of cut~\#3 illustrated in the inset. The data were taken by He lamp ($h\nu$=21.2~eV) at 50K and divided by resolution-convolved Fermi Dirac Function.
 (\textbf{m}) EDCs from -0.4~\AA$^{-1}$ to 0.4~\AA$^{-1}$ for the data in panel~(l). The red open circles were determined by the local maxima or shoulders.
(\textbf{n}) Parabolic fitting of the $\alpha$ band dispersion and linear fitting to extract the Fermi velocity $v_{\mathrm{F}}$, along with a schematic illustration of Hund-assisted electron correlations and additional many-body renormalization effects on the few-meV energy scale.
}
	\label{fig:figure212}
\end{figure*}

In typical $f$-electron heavy-fermion systems like CeCoIn$_5$, ARPES measurements have shown that the quasiparticle spectral weight of the flat Ce~4$f$ band near the Fermi level increases logarithmically with decreasing temperature, without saturation down to 17~K~\cite{PhysRevB.96.045107}. 
As the hybridization between the flat $f$-electron band and the dispersive conduction band strengthens, the Fermi momentum $k_\mathrm{F}$'s varies with temperature, leading to a gradual reconstruction of the Fermi surface~\cite{PhysRevB.96.045107}. To examine whether similar phenomena occur in a 3$d$-electron heavy fermion system, we performed temperature-dependent ARPES measurements to track the evolution of the flat band in LiV$_2$O$_4$.
As temperature decreases, the flat band of LiV$_2$O$_4$ becomes increasingly prominent [Figs.~4(a)-4(h)].
In the momentum-integrated EDCs, the flat band peak sharpens at lower temperatures, indicating enhanced coherence. However, the spectral weight of this peak spans a broad energy range exceeding 50~meV, far larger than the energy scale of heavy quasiparticles expected from $k_\mathrm{B}T^*\approx$~2.6~meV. 
In contrast to CeCoIn$_5$, the flat band weight of $\alpha$ in LiV$_2$O$_4$ does not follow a logarithmic temperature dependence [Fig.~4(j)].
Instead, the flat band intensity is already present at 100~K and gradually saturates below 50~K, exhibiting no obvious correlation with the crossover temperature $T^*$$\approx$~30K. This saturation may reflect the completion of orbital screening characteristic of Hund’s metals, which can take place at a much higher temperature than the screening of spins \cite{deng2019signatures}.
Furthermore, as shown in Fig.~4(k), both the width and position of the MDC peaks at the Fermi level are nearly temperature-independent, in contrast to the $c$-$f$ hybridization in CeCoIn$_5$, indicating stable orbital occupancy of the $a_{1g}$ states. Consistently, orbital-selective NMR measurements have suggested minimal temperature dependence in orbital populations~\cite{shimizu2012orbital}.
The saturation of flat band intensity above 50~K and the lack of temperature-driven $k_\mathrm{F}$ variation distinguish the emergence of flat band in LiV$_2$O$_4$ from $c$-$f$ hybridization in typical heavy-fermion systems. While the flat band in LiV$_2$O$_4$ reflects strong correlations, the additional formation of heavy quasiparticles associated with this flat band likely occurs at a much lower energy scales below $T^*$.

An electron-like dispersion of the $\alpha$ band, with a bandwidth of approximately 25~meV, is resolved near $E_\mathrm{F}$ [Figs.~4(l) and 4(m), and Appendix C]. This bandwidth is strongly renormalized compared to the corresponding DFT prediction [Fig.~2(g)].

Such a flat band is missing in previous angle-integrated photoemission spectroscopy studies~\cite{shimoyamada2006heavy}, possibly due to potential surface or crystal quality issues, a situation similar to early studies on CeCu$_2$Si$_2$~\cite{reinert2001temperature,wu2021revealing}.  
Using the band bottom and Fermi crossing points, a parabolic estimation gives an effective mass of approximately 19~$m_\mathrm{e}$ [Fig.~4(n)], which reflects the renormalization of the entire bandwidth by Hund-assisted electron correlation. 
However, closer to the Fermi level, the band dispersion is noticeably flatter, deviating from a simple parabolic dispersion. This feature is consistently observed at various temperatures (Appendix D) and signals additional renormalization on a few-meV energy scale. The Fermi velocity is estimated to be $v_\mathrm{F}$~=~0.034~eV$\cdot$\AA, which is 3\% of that of the $\beta$ band.
By comparing the $v_\mathrm{F}$ with the slope of parabolic estimation at $E_\mathrm{F}$, the additional renormalization further enhances the effective mass to at least 93~$m_\mathrm{e}$. 
This few-meV renormalization scale is consistent with the expected energy scale of heavy quasiparticles ($k_\mathrm{B}T^*\approx$~2.6~meV), and the enhanced effective mass is comparable to the value of 180~$m_\mathrm{e}$ inferred from specific heat measurements~\cite{JOHNSTON200021}. The remaining discrepancy, approximately a factor of 2, likely arises from the continued development of low-energy renormalization at lower temperatures and the limited energy resolution of our measurements, which may not fully capture the small energy scale involved. An accurate determination of the effective mass and full characterization of the quasiparticle dispersion will require sub-meV energy resolution at temperatures below 2~K, when the heavy Fermi liquid state is fully developed~\cite{urano2000liv}.

$3.~~Discussions.$~~The entire bandwidth of $\alpha$ is substantially renormalized with respect to DFT calculations [Fig.~2(g)]. Such bandwidth narrowing and mass enhancement are characteristic of strong electron correlationsand are commonly observed in systems near Mott transitions. For example, in NiS$_{2-x}$Se$_x$, increasing correlation strength gradually reduces the bandwidth as the system approaches the Mott transition~\cite{PhysRevLett.112.087603}. Analogously, in LiV$_2$O$_4$, the strong correlations within the $a_{1g}$ orbital may push it close to an orbital-selective Mott-Hubbard transition.
Although the coherent weight of the $\alpha$ band does not correspond to the Kondo screening in typical heavy fermion systems, its saturation may reflect the full screening of orbital degrees of freedom, as expected in the Hund’s metal scenario \cite{deng2019signatures}. The well developed coherent dispersion of $\alpha$ band establishes the prerequisite for the formation of lower-energy heavy quasiparticles.
 
The additional renormalization at $E_\mathrm{F}$ suggests coupling with excitations on the few-meV energy scale.
Nuclear magnetic resonance (NMR) studies indicate antiferromagnetic correlations between V atoms~\cite{shimizu2012orbital}, while inelastic neutron scattering experiments have detected short-range spin fluctuations with characteristic energies from 0.2 to 0.8~meV~\cite{lee2001spin}. In the geometrically frustrated pyrochlore sublattice formed by V atoms, such low-energy spin excitations are expected to remain dynamic and can couple to electrons near the Fermi level. This coupling may have induced the observed additional renormalization on a few-meV energy scales, enhancing the effective mass beyond the renormalization of the entire bandwidth by electron-electron interactions and Hund's coupling.

In summary, our study provides the first direct observation of the three-dimensional electronic structure of LiV$_2$O$_4$. 
We identify a flat, heavy $\alpha$ band associated with the $a_{1g}$ orbital, whose strong mass renormalization arises from cooperative effects of Hund’s coupling and electronic correlations. 
Furthermore, we uncover higher-order electron-boson coupling in the itinerant $\beta$ band, revealing further many-body interactions that should be incorporated into theoretical models.
In addition to the renormalization of the $a_{1g}$ orbital over the entire bandwidth, possible low-energy interactions with excitations induced by geometric frustration contribute to the exceptionally large effective mass inferred from thermodynamic measurements.
These findings advance our understanding of heavy fermion physics in 3$d$ transition-metal systems and lay the groundwork for future exploration of exotic quantum phases in correlated oxides.

$Note$: During the revision of our manuscript, we became aware of the preprint of an independent study by Dongjin Oh, \textit{et al.}~\cite{oh2025hundflatbandfrustrated}, which also reported measurements of the three-dimensional electronic structure and flat band of LiV$_2$O$_4$.

\section*{Acknowledgements}
We gratefully acknowledge the valuable discussion with Dr. P. Y. Zheng, Dr. Z. H. Yuan and Dr. Y. L. Wang, the experimental support of Dr. Z. T. Liu, Dr. Z. C. Jiang, Dr. Marta Zonno and Dr. Sergey Gorovikov.
This work is supported in part by the National Key R\&D Program of the MOST of China (2023YFA1406300), the National Science Foundation of China under the grant Nos. 12274085, 12422404, 92477206, the New Cornerstone Science Foundation, the Innovation Program for Quantum Science and Technology (Grant No. 2021ZD0302803), and Shanghai Municipal Science and Technology Major Project (Grant No.2019SHZDZX01). 
The ARPES measurements used Beamlines 09U and 03U of the Shanghai Synchrotron Radiation Facility and Beamline QMSC of Canadian Light source. The Beamline 03U is supported by the ME2 project under contract no. 11227902 from National Natural Science Foundation of China.
%Natural Science Foundation of China (Grants No. 11888101, 12074074, and 11790312), the National Key R\&D Program of the MOST of China (Grants No. 2017YFA0303004 and 2016YFA0300200), Project supported by Shanghai Municipal Science and Technology Major Project (Grant No. 2019SHZDZX01), and Shanghai Rising-Star Program (20QA1401400).
	
%[Competing Interests] The authors declare that they have no competing financial interests.
    
\appendix

\section{\label{I}The LiV$_2$O$_4$ thin film growth method and Samples characterization}

\begin{figure}[htbp]
\renewcommand*{\thefigure}{A1}
\includegraphics[width=71mm]{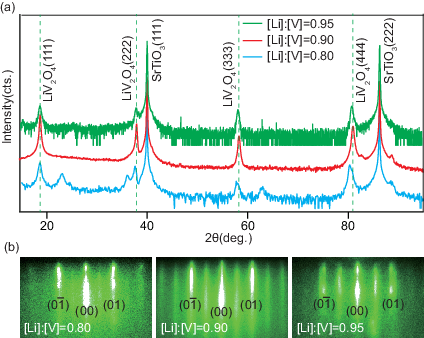}
\caption{
(\textbf{a}) XRD patterns of LiV$_2$O$_4$ thin films grown using targets with different Li:V ratios. The dashed lines indicate the characteristic diffraction peaks of bulk single-crystal LiV$_2$O$_4$ along the [111] orientation. 
(\textbf{b}) The RHEED pattern of LiV$_2$O$_4$ thin film grown by different composition targets.
}
\label{band evolution}
\end{figure}

To achieve optimal film quality, we systematically varied the Li:V ratio in ceramic LiV$_2$O$_4$ targets used for PLD. Targets with a Li:V ratio of 0.8 exhibited secondary-phase diffraction peaks [Fig.~A1(a)] and crescent-shaped RHEED patterns [Fig.~A1(b)], indicating chemical inhomogeneity and rough surface morphology. Increasing the Li content to 0.95 suppressed the secondary phases [Fig.~A1(a)] but resulted in degraded crystallinity, as evidenced by the emergence of three-dimensional diffraction features in RHEED [Fig.~A1(b)]. The optimal target composition was found to be Li:V = 0.9, which yielded single-phase films exhibiting sharp X-ray diffraction (XRD) peaks [Fig.~A1(a)] and two-dimensional RHEED streaks [Fig.~A1(b)]. Furthermore, XRD patterns of films grown with the optimal Li:V = 0.9 target composition show excellent agreement with bulk single-crystal LiV$_2$O$_4$ along the [111] orientation. This confirms that the films are fully relaxed and retain the same crystallographic structure as the bulk material.

\section{\label{II}Polarization-Dependent ARPES Measurement}
\begin{figure}[htbp]
    \renewcommand*{\thefigure}{A2}
    \includegraphics[width=87mm]{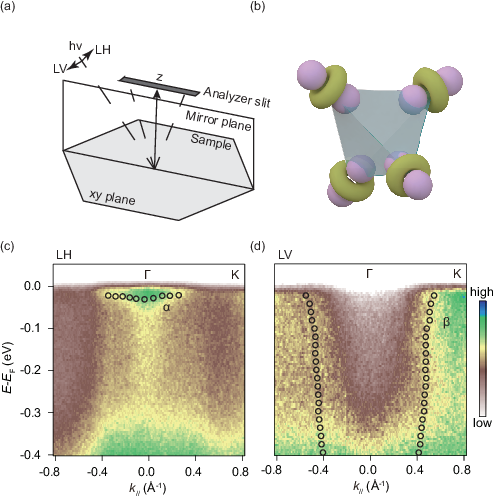}
    \caption{
(\textbf{a}) Experimental setup for polarization-dependent ARPES. For the  linearly horizontal (LH) and  linearly vertical (LV) experimental geometry, the electric field direction of the incident photons is parallel (or perpendicular) to the mirror plane defined by the analyzer slit and the sample surface normal. 
(\textbf{b}) The spatial orientation of the $a_{1g}$ orbital in a tetrahedron formed by four vanadium atoms at its vertices.
(\textbf{c})-(\textbf{d}) The photoemission spectra along the $\Gamma$-K direction were obtained using LH or LV polarized light. The experimental setup is depicted in panel (a). The $\alpha$ and $\beta$ bands are indicated by circles determined from the local maxima of MDCs and EDCs.
}
    \label{fig:polarization}
\end{figure}
\begin{figure}[htbp]
    \renewcommand*{\thefigure}{A3}
    \includegraphics[width=87mm]{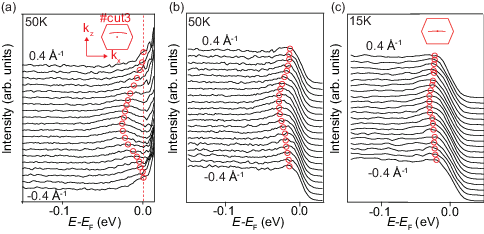}
    \caption{(\textbf{a}) The EDCs extracted from -0.4~\AA$^{-1}$ to 0.4~\AA$^{-1}$ for the data in Fig.~4(l). The red open circles were determined by the local maxima. EDC has been divided by the Fermi–Dirac function. (\textbf{b}) The EDCs in panel (a) are not divided by the Fermi–Dirac distribution. (\textbf{c}) The EDCs measured at a photon energy of 101~eV were acquired at the momentum position indicated by the inset in the Brillouin zone.}
    \label{fig:arpes_cuts}
\end{figure}

\setlength{\textfloatsep}{5pt plus 2pt minus 2pt}

\begin{figure}[htbp]
    \renewcommand*{\thefigure}{A4}
    \includegraphics[width=86mm]{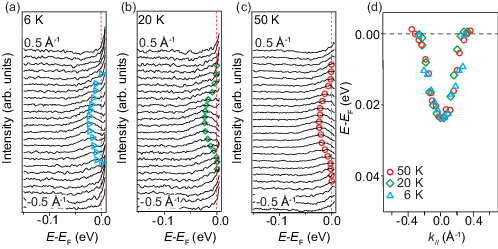}
    \caption{Temperature-dependent dispersion of the $\alpha$ band. (\textbf{a}-\textbf{c}) EDC-divided spectra at 6~K, 20~K and 50~K, respectively. The dispersions of the $\alpha$ band were extracted from EDC peak positions at each temperature. (\textbf{d}) Overlaid electron-like dispersions of the $\alpha$ band at the three temperatures.}
    \label{fig:arpes_cuts}
\end{figure}

Polarization-dependent ARPES measurements show that the flat band near the Fermi level is visible under linear horizontal (LH) polarization but significantly suppressed under linear vertical (LV) polarization~[Figs.~A2(c) and (d)]. Due to the complex spatial orientation of the V $a_{1g}$ orbitals~[Fig.~A2(b)], direct analysis of the matrix element effects based on simple symmetry arguments is challenging. Nevertheless, the strong polarization sensitivity observed in our ARPES data, combined with theoretical calculations, suggests that the $\alpha$ band predominantly originates from the $a_{1g}$ orbital.

\begin{figure}[t]
  \centering
  \renewcommand*{\thefigure}{A5}
  \includegraphics[width=53mm]{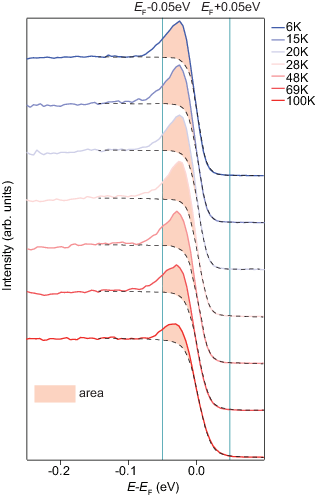}
  \caption{The process of the flat band weight extraction from the EDCs of Fig.~4(i). 
  The dashed curves indicate the incoherent background fitted by resolution-convolved Fermi--Dirac function. }
  \label{fig:arpes_cuts}
\end{figure}

\section{\label{III}The Electron-like $\alpha$ Flat Band}

To confirm whether the flat $\alpha$ band remains electron-like across different photon energies, we conducted ARPES measurements at multiple photon energies and along different momentum cuts. In Fig.~A3(a), the spectra are divided by the Fermi–Dirac distribution to enhance visibility near the Fermi level, revealing the flat $\alpha$ band. Fig.~A3(b) shows the corresponding spectra without Fermi–Dirac distribution division. Here, the $\alpha$ band remains visible, and its peak position in the EDCs is still located below the Fermi level, confirming an electron-like dispersion. Furthermore, Fig.~A3(c) shows that the band structure measured at 101 eV without Fermi-Dirac division also exhibits the electron-like dispersion of the $\alpha$ band. These observations confirm that the $\alpha$ band is an intrinsic feature and retains its electron-like dispersion under varying experimental conditions.

\section{\label{IV}Temperature-Independent Dispersion of the $\alpha$ Band}

To examine the temperature dependence of the $\alpha$ band dispersion, we conducted measurements at 6~K~[Fig.~A4(a)], 20~K~[Fig.~A4(b)], and 50~K~[Fig.~A4(c)]. At each temperature, the raw spectra were divided by the corresponding Fermi–Dirac distribution to enhance the visibility of states near the Fermi level. The $\alpha$ band dispersions were extracted by tracking the peak positions in the EDCs, and all three dispersions were plotted together for direct comparison. As shown in Fig.~A4(d), the extracted electron-like dispersions of the $\alpha$ band exhibit negligible change across the measured temperature range. In the main text, we present analysis performed on the data taken at the elevated temperature of 50~K, because the thermal excitations at this temperature enable the detection of unoccupied states slightly above $E_\mathrm{F}$ and help to more clearly resolve the signatures of additional renormalization in the $\alpha$ band near $E_\mathrm{F}$.

\section{\label{V} Extraction of the flat band weight}

We extract the flat-band weight from the area of the EDC peak in Fig~4(i), as illustrated in Fig.~A5. To determine and remove the incoherent background, the EDC measured at 100~K is fitted with a resolution-convolved Fermi–Dirac function. The background curves are then generated for each temperature. The flat-band weight is defined as the integrated spectral weight within the window [$E_\mathrm{F}$ - 0.05~eV, $E_\mathrm{F}$ + 0.05~eV] after subtracting the background curve.

%\FloatBarrier
	
\bibliographystyle{apsrev4-2}
\bibliography{sample.bib}

\end{document}